\documentclass[%
reprint,
amsmath,amssymb,
aps,
showkeys,
]{revtex4-1}

\usepackage{graphicx}
\usepackage{dcolumn}
\usepackage{bm}
\usepackage{ dsfont }

\newcommand{\niftyt}{\textsc{NIFTy\,3}}

\begin{document}
	
	\preprint{APS/123-SDE}
	
	\title{Field dynamics inference via spectral density estimation}
	
	\author{Philipp Frank}
	
    \author{Theo Steininger}
    
	\author{Torsten A. En\ss lin}%
	\affiliation{%
		Max-Planck Institut f{\"u}r Astrophysik, Karl-Schwarzschild-Str. 1, 85748, Garching, Germany
	}%
	\affiliation{%
		Ludwig-Maximilians-Universit{\"a}t M{\"u}nchen, Geschwister-Scholl-Platz 1, 80539, M{\"u}nchen, Germany
	}%

	\date{\today}
	
	\begin{abstract}
		Stochastic differential equations (SDEs) are of utmost importance in various
		scientific and industrial areas. They are the natural description of
		dynamical processes whose precise equations of motion are either not known
		or too expensive to solve, e.g., when modeling Brownian motion. In some cases,
		the equations governing the dynamics of a physical system on macroscopic
		scales occur to be unknown since they typically cannot be deduced from
		general principles. In this work, we describe how the underlying laws of a stochastic process can be approximated by the spectral density of the corresponding process. Furthermore, we show how the density can be inferred from possibly very noisy and incomplete
		measurements of the dynamical field. Generally, inverse problems like these
		can be tackled with the help of \emph{Information Field Theory} (IFT). For now, we restrict to
		\emph{linear} and \emph{autonomous} processes. Though, this is a non-conceptual limitation that may be omitted in future work. To demonstrate its
		applicability we employ our reconstruction algorithm on a time-series and spatio-temporal processes.
		
	\end{abstract}
	
	\maketitle
	
	
\section{Introduction}
Stochastic differential equations (SDEs) play an important role in a variety of scientific fields \cite{henderson2006stochastic} and industrial designs \cite{Iacus:2008:SIS:1386239}. SDEs are a flexible tool to model dynamical processes and traditionally serve as a useful prior for the inference of dynamical fields. In recent decades the inverse problem gained considerable attention as well. The goal of the inverse problem is to infer the dynamical properties of the system from observations of an evolving field. In order to tackle this problem a variety of methods have been proposed in the past \cite{2017arXiv170310112K,Roweis:1999:URL:309394.309396}. These range from parametric methods which aim to parametrize the process either in the temporal \cite{RePEc:arx:papers:1603.05700} or the frequency domain \cite{Spangl2013} to non-parametric and Bayesian methods \cite{DBLP:conf/nips/RuttorBO13,bretthorst1988bayesian,Macaro2014}. 

In a number of physical disciplines (e.g., astrophysics or plasma-physics) the dynamical quantities of interest may evolve not only in time but are also extended in space. To infer the underlying spatio-temporal  dynamical process, a variety of methods have been proposed. These include kernel-based methods \cite{Knauf2016}, approaches using spatio-temporal Kriging \cite{JTSA:JTSA12245} as well as Bayesian non-parametric approaches for multi-dimensional spatial evolution \cite{doi:10.1093/biomet/asp066}. Some methods aim to tackle this problem assuming separability of the spatial and the temporal evolution which helps to simplify the inference problem. In many physical applications, however, separability cannot be assumed due to the entanglement of structures in space and time.

For autonomous processes, the dynamics is fully determined in terms of the spectral density of the dynamical field. In this work we will use this property to model linear, non-separable, SDEs. Furthermore, the proposed method is able to reconstruct the spectral density also for noisy and masked observations of the dynamical field using Bayesian inference.

Physical fields usually can be defined in a corresponding configuration space and linear differential equations can be described as linear operators acting on field vectors in this space. We model the problem with the help of probabilities over the elements of this configuration space directly to reflect the mentioned properties of the field. To this end, we rely on the language of information field theory (IFT) \cite{2013AIPC.1553..184E,2009PhRvD..80j5005E} which extends information theory to fields. The restriction of this work to spatially and temporal autonomous differential operators enables us to describe those in terms of fields over the joint Fourier spaces and to derive posterior distributions for the corresponding fields directly. Due to the fact that the resulting equations are often not analytically traceable, the computations ultimately are performed on a finite grid on a computer. However, as IFT is independent of the chosen discretization, one can always choose a representation which is convenient for the problem at hand. In this work, the discretization is achieved using the software package \niftyt \ (Numerical Information Field Theory) \cite{2017arXiv170801073S} which allows building algorithms using a field theoretical language, while the framework takes care of the underlying discretization.
In order to introduce the notation and field theoretical language used throughout this paper, we continue with a brief introduction to IFT.

\subsection{Introduction to IFT and notation}\label{sec:ift}
IFT is a statistical field theory which allows doing probabilistic calculations for fields (in the physical sense), defined over continuous space (or space-time). It is already used in various inference tasks such as astrophysical imaging \cite{2014AIPC.1636...49E}, component separation \cite{2017arXiv170502344K}, numerical simulations of dynamical systems \cite{2013PhRvE..87a3308E} and others. IFT has also previously been applied in the context of dynamical system inference \cite{2016PhRvE..94a2132P}.
The theory enables the user to work directly in the field configuration space, which for many applications involving fields is the natural space to define problems.

In order to define probabilities for fields we introduce a scalar product for fields ($\phi(x)$, $\psi(x)$ with $x \in \Omega = \mathds{R}^n$) as
\begin{equation}
	\phi ^\dagger \psi = \int\limits_{\Omega} \phi^*(x) \psi(x) \ dx \ ,
\end{equation}
where $^*$ denotes complex conjugation. Note that throughout this paper we restrict ourselves to scalar fields, for simplicity, although a generalization to vector fields is possible.
As an example, a Gaussian distribution for a field can be written as
\begin{equation}\label{eq:gauss}
	\mathcal{P}(\phi) = \mathcal{G}(\phi-m,\Phi) = \frac{1}{\sqrt{\left| 2 \pi \Phi \right|} } \exp\left\lbrace - \frac{1}{2} \phi ^\dagger \Phi^{-1} \phi \right\rbrace  \ ,
\end{equation}
where $m$ denotes the mean and $\Phi$ is a linear, self-adjoint and strictly positive operator which maps from the field configuration space to itself. In other words, it is the continuous version of a covariance. Note that $\Phi$ is sometimes referred to as a covariance matrix, although it is actually a continuous operator.
$\left| \Phi\right| $ denotes the functional determinant of the operator $\Phi$. The exponential factor in Eq.\,\ref{eq:gauss} reads
\begin{equation}
	\phi ^\dagger \Phi^{-1} \phi = \phi^*_x \Phi^{-1}_{xy} \phi_y = \int dx \int dy  \ \phi^*(x) \ \Phi^{-1}(x,y) \ \phi(y) \  ,
\end{equation}
where we also introduced the continuous version of the Einstein sum convention.

Although it is natural to formulate many physical problems in a field theoretical language, the data that we observe is always finite. Therefore, we need to translate between the field configuration space (called signal space) and the so called data space. To illustrate this procedure consider the following measurement scenario
\begin{equation}\label{eq:lindata}
d_u =  R_u[s] + n_u \ ,
\end{equation}
where $R$ denotes a projection operator which projects from the infinite dimensional configuration space of $s$ to a finite set of  $M$ measurement points $d_u$ (with $u \in \{1,..,M\}$) and $n$ is the measurement noise. Using this data model and Bayes theorem, the posterior of $s$ given $d$ reads
\begin{equation}
	\mathcal{P}(s|d) \propto \int \mathcal{D}n \ \mathcal{P}(d|s,n) \mathcal{P}(s) \mathcal{P}(n) \ .
\end{equation}
If we assume $n$ and $s$ to be independently Gaussian distributed with zero mean and covariances $N$ and $S$, respectively, and also assume $R$ to be a linear operator the posterior is also a Gaussian. I.e.
\begin{equation}
	\mathcal{P}(s|d) = \mathcal{G}(s-m,D) \ ,
\end{equation}
with mean
\begin{equation}\label{eq:wf}
	m = D j = (S^{-1} + R^\dagger N^{-1} R)^{-1} R^\dagger N^{-1} d \ ,
\end{equation}
and covariance $D=(S^{-1} + R^\dagger N^{-1} R)^{-1}$. These equations resemble the famous Wiener filter equations \cite{wiener_extrapolation_1950} which is the optimal linear filter, applied to a field theoretical setting.

In a non-linear setting, however, the exact form of the posterior or corresponding expectation values are often not analytically traceable. Therefore, in this work we rely on maximum a posteriori (MAP) estimates which we obtain by minimizing the so called information Hamiltonian defined as
\begin{equation}\label{eq:hamilton}
	\mathcal{H}(x) = - \log\left( \mathcal{P}(x)\right) \ ,
\end{equation}
where $\log(x)$ is the natural logarithm.
In addition, the second derivatives of the Hamiltonian give rise to the Laplace approximation of the uncertainty maps, as we will discuss in appendix \ref{ap:grad}.

To motivate the application of IFT to the inference of dynamical systems we note that linear differential equations can be rewritten as linear (differential) operators acting on a field of interest. As we will point out in the next section, these operators serve as a building block for the covariance function in the context of SDEs. Specifically, we draw the connection to the spectral density of the field which is defined as the diagonal of the covariance operator in harmonic space.

\subsection{Structure of this work}
The rest of this work structures as follows:
In section \ref{sec:sde} we briefly outline how SDEs are connected to the spectral density of a random process. Consequently, in section \ref{sec:spect} we describe the key properties of our inference method of spectral densities from field realizations as well as noisy measurements thereof. In section \ref{sec:application} we apply our method to different mock data examples including one- and two-dimensional examples. Finally, in section \ref{sec:conclusion}
we conclude the paper with a short summary as well as a small outlook to possible applications and further projects.

\section{From a SDE to the spectral density}\label{sec:sde}
In this section we outline how the properties of a linear SDE are encoded in the spectral density of a spatio-temporal dynamical process. Furthermore, we introduce the key assumptions that are necessary to ensure that all relevant information is encoded in the density.

A suitable starting point is a linear SDE of the form:
\begin{equation}\label{eq:linmodel}
(\mathcal{L}\phi)(\mathbf{x},t) = \xi(\mathbf{x},t) \ ,
\end{equation}
where $\mathcal{L}$ is a linear (differential) operator acting on a field $\phi$, and $\xi$ is a random process. $\mathbf{x} \in \mathds{R}^d$ and $t \in \mathds{R}$ denote the spatial and temporal coordinates, respectively. Note that the distinction between space and time is for convenience only, i.e. all formulas treat space and time on the same footage which means that the analysis is also valid for a general multi-dimensional process.

Assuming $\xi$ to be Gaussian distributed with a covariance matrix $\Theta$ and using Eq.\,\ref{eq:linmodel} yields the probability distribution for $\phi$:
\begin{equation}\label{eq:likeli}
\mathcal{P}(\phi|\mathcal{L},\Theta) = \int  \mathcal{D}\xi \  \mathcal{P}(\phi|\mathcal{L},\xi) \ \mathcal{G}(\xi,\Theta) = \mathcal{G}(\phi,\Phi) \ ,
\end{equation}
where
\begin{equation}\label{eq:Sk}
\Phi = (\mathcal{L}^{\dagger}\Theta^{-1} \mathcal{L})^{-1} \ .
\end{equation}
As we can see, all relevant information concerning the statistical properties as well as the dynamic evolution is encoded in the correlation matrix $\Phi$ if the underlying process is linear.
 
Assuming $\mathcal{L}$ to be local and homogeneous in both, space and time, implies that the operator can be written as
\begin{equation}\label{eq:Lx}
\mathcal{L}_{xx'} = \delta^{(d+1)}(x-x')  \ g(\partial_t,\partial_{\mathbf{x}}) \ ,
\end{equation}
where we introduced the space-time vector $x=(t,\mathbf{x})$ and the differential operator encoding function $g$.
Fourier transforming Eq.\,\ref{eq:Lx} yields:
\begin{equation}
\mathcal{L}_{kk'} = (2 \pi)^{d+1} \delta^{(d+1)}(k-k') \ f(k) \ ,
\end{equation}
where $k=(\omega,\mathbf{k})$ denotes the coordinates in harmonic space and $f(k) = g(i \ \omega,i \ \mathbf{k})$ is a complex scalar field, with $i$ being the imaginary unit. Note that if the differential equation is real, then $f^*(\omega,\mathbf{k}) = f(-\omega,-\mathbf{k})$ and therefore $\mathcal{L}$ is Hermitian.

Assuming further that also $\Theta$ is diagonal in harmonic space with the spectral density $P_{\xi}(k)$, Eq.\,\ref{eq:Sk} can be rewritten as
\begin{align}
	\Phi_{kk'} &= (2 \pi)^{d+1} \delta^{(d+1)}(k-k') \frac{P_{\xi}(k)}{\left| f(k) \right|^2 }
	\notag\\ & =:  (2 \pi)^{d+1} \delta^{(d+1)}(k-k') P_{\phi}(k)\ ,
\end{align}
where we defined $P_{\phi}(k)$, the spectral density of $\phi$.

As we can see, $P_{\phi}$ encodes the properties of a SDE up to the complex phase of $f$. Therefore, we seek to find a way to infer it from observations of the field $\phi$.
In order to derive the posterior distribution of the spectrum, in the following, we propose a way to model the key features of $P_{\phi}$.

\section{Spectral density inference}\label{sec:spect}
In order to model $P_{\phi}$ we notice that if $f$ and $P_{\xi}$ are continuous and smooth functions of their arguments, then $P_{\phi}$ is a rational and positive function. We therefore model the spectral density as
\begin{equation}
	P_{\phi}(k) = \exp\left[ \tau(k)+\tan\left( \delta(k)\right) \right] \ .
\end{equation}
The idea behind this definition is that we want to reduce $P_{\phi}$ to its two key properties: either $P_{\phi}$ is a smooth, positive function of $k$, which we model by $\exp(\tau)$, or it diverges as
\begin{equation}
	\left| f(k) \right|^2 \rightarrow 0 \ ,
\end{equation}
which is modeled by $\exp(\tan(\delta))$.
We therefore define suitable prior distributions for $\tau$ and $\delta$ which aim to support these features.

\subsection{$\tau$-Prior}
In order to constrain $\tau$ to be a smooth function of its arguments, we impose a smoothness prior on $\tau$ (see e.g., \cite{2013PhRvE..87c2136O}).
To get an idea about smoothness and the corresponding prior consider the following 1D example:
Suppose $P_\phi$ models a power-law, i.e.
\begin{equation}
	P_\phi(y) = y^\alpha \ ,
\end{equation}
with $y,\alpha \in \mathds{R}$ and assume for the moment $\delta=0$.
Then
\begin{equation}
	\tau = \log\left(P_\phi(y)\right)  = \alpha \log\left( y\right) \ ,
\end{equation}
is linear in $\log(y)$ which implies that the second logarithmic derivative vanishes. 
We therefore built our prior such that it minimizes the curvature of $\tau$ on a logarithmic scale.
This reads
\begin{equation}
	\mathcal{P}(\tau)=\mathcal{G}(\tau,T_\sigma) \ ,
\end{equation}
with $T_\sigma$ such that
\begin{equation}
	\tau^\dagger T_\sigma^{-1} \tau = \frac{1}{\sigma^2} \int d\left( \log\left( y\right)\right)  \left|\frac{\partial^2 \tau(y)}{\partial \log\left( y\right)^2}\right|^2 \ , \ \sigma \in \mathds{R} \ ,
\end{equation}
where $\sigma$ is an overall hyper-parameter controlling the degree of smoothness one wants to impose on $\tau$.

In higher dimensions, this constraint has to be imposed for all quadratic, logarithmic variations of the field simultaneously. 
A derivation of the exact form as well as a short discussion can be found in appendix \ref{ap:smooth}. Furthermore, in some applications it is necessary to impose smoothness also for negative $y$. To do so, we extend this prior using the complex logarithm, which is also defined on a negative scale. Details of this approach as well as the treatment of the special point $y=0$, are discussed in appendix \ref{ap:comlog}.

\subsection{$\delta$-Prior}
The prior distribution for $\delta$ is constructed in a way such that $\delta$ allows for a transition of the spectral density from smooth to divergent regions. We therefore also impose a smoothness prior on $\delta$ which implies that for small $\delta$, where
\begin{equation}
	\tan(\delta) \approx \delta \ ,
\end{equation}
the spectrum remains smooth.
However, as $\delta$ approaches $\pm \pi/2$, small changes in $\delta$ result in large, abrupt changes of the spectrum.

The full prior reads
\begin{equation}
	\mathcal{P}(\delta) \propto \mathcal{G}(\delta,T_\mu) \ \mathcal{G}(\delta,\nu^2 \mathds{1}) \ , \ \delta \in \left[-b,b\right]  \ , \ \mu, \nu \in \mathds{R} \ ,
\end{equation}
where we also included a term to the prior that punishes larger values of $\delta$. This ensures that $\delta$ remains zero in regions where the data does not support a divergence. Note that we restrict the support of $\delta$ to $b = (\pi / 2 - \epsilon)$, where $\epsilon$ serves as a ``high-energy'' cutoff in order to avoid infinities during reconstruction. Since the length-scale of $\delta$ is $\pi/2$ we note that $\nu \approx \pi/2$ is a reasonable choice.

\subsection{Perfect data Posterior}\label{sec:post1}
Using the priors and the likelihood, defined in Eq.\,\ref{eq:likeli}, we can immediately write down the posterior distribution
\begin{equation}
	\mathcal{P}(\tau,\delta | \phi) \propto \mathcal{G}(\phi,\Phi) \ \mathcal{P}(\tau) \ \mathcal{P}(\delta)
\end{equation}
and the corresponding information Hamiltonian
\begin{align}\label{eq:h1}
	& \mathcal{H}(\tau,\delta | \phi) =- \log\left( \mathcal{P}(\tau,\delta | \phi) \right) \notag\\
	 &= \frac{1}{2} \left[ \phi^\dagger \Phi^{-1} \phi+\log\left(\left| \Phi\right|   \right) +  \tau^\dagger T_\sigma^{-1} \tau + \delta^\dagger\left(  T_\mu^{-1}+ \nu^{-2} \right)  \delta  \right] \notag\\ &+ H_0 \ ,
\end{align}
where $H_0$ is a constant that is independent of $\tau$ and $\delta$. Minimizing this Hamiltonian with respect to $\tau$ and $\delta$ leads to their maximum a posteriori estimates, given perfect data on the realization of $\phi$.

\subsection{Noisy data Posterior}\label{sec:post2}
In reality, we are usually only able to retrieve noisy measurement data, $d$, of $\phi$. We therefore seek to find a way to infer the spectral density from noisy measurements rather than from $\phi$ itself. Using the notation introduced in section \ref{sec:ift} and the data model (Eq. \ref{eq:lindata}), the joint distribution reads
\begin{equation}\label{eq:joint}
	\mathcal{P}(d,\phi,\tau,\delta) = \mathcal{G}(d-R\phi,N) \mathcal{G}(\phi,\Phi)\mathcal{P}(\tau)\mathcal{P}(\delta) \ .
\end{equation}
Depending on the measurement process, in particular the form of $R$, the optimal way to proceed may differ significantly. As this is a general issue concerning Bayesian inference and well discussed in literature, we want to focus the discussion on our definition of the spectral density rather than all possible ways of inference.

However, we note that there exist in principle two different approaches to reconstruction in this case. One way is to minimize the information Hamiltonian corresponding to Eq.\,\ref{eq:joint}, with respect to all quantities of interest ($\phi$, $\tau$, $\delta$), to obtain a maximum a posterior solution. In cases of high quality data, this is a good way to proceed.

However, in cases of high measurement uncertainty, the high frequency modes of $\phi$ are suppressed in the reconstruction as they are indistinguishable from the noise. Due to the fact that the natural domain of ($\tau$, $\delta$) is the harmonic domain, the lack of high frequency modes restricts a reliable reconstruction of $\tau$ and $\delta$ to low frequencies.

A possible way to resolve this issue is to marginalize out $\phi$ in Eq.\,\ref{eq:joint}. For a linear $R$ marginalization is obtained analytically and the marginal information Hamiltonian is given by
\begin{align}\label{eq:marginalhamilton}
	\mathcal{H}(\tau,\delta|d) &= \frac{1}{2}\left( \log\left( \frac{\left| \Phi \right|}{\left| D\right|} \right) - j^\dagger D j +\tau^\dagger T_\sigma^{-1} \tau + \delta^\dagger T_\mu^{-1} \delta \right) \notag\\ 
	&+ \frac{1}{2 \nu^2} \delta^\dagger  \delta   +H_0 \ ,
\end{align}
with
\begin{equation}\label{eq:dj}
	D = \left(R^\dagger N^{-1} R + \Phi^{-1} \right)^{-1} , \ j = R^\dagger N^{-1} d \ ,
\end{equation}
analogous to Eq.\,\ref{eq:wf}.
Minimizing Eq.\,\ref{eq:marginalhamilton} leads to the posterior estimates which we will call $\bar{\tau}$ and $\bar{\delta}$ in the following.

In the spirit of the empirical Bayes approach \cite{Bishop:1995:NNP:525960}, where we treat these estimates as the true values irrespective of their corresponding uncertainties, the approximate posterior of $\phi$ reads
\begin{equation}\label{eq:empbayes}
	\mathcal{P}(\phi|d,\bar{\tau},\bar{\delta}) = \mathcal{G}(\phi-\bar{D}j,\bar{D}) \ ,
\end{equation}
where $\bar{D}$ denotes the information propagator, evaluated at $\bar{\tau}$, $\bar{\delta}$.

Since now all ingredients that are necessary for inference are available, consistency tests as well as mock data applications are presented in the next section.

\begin{figure}[h]
	\includegraphics[scale=0.6, angle=0]{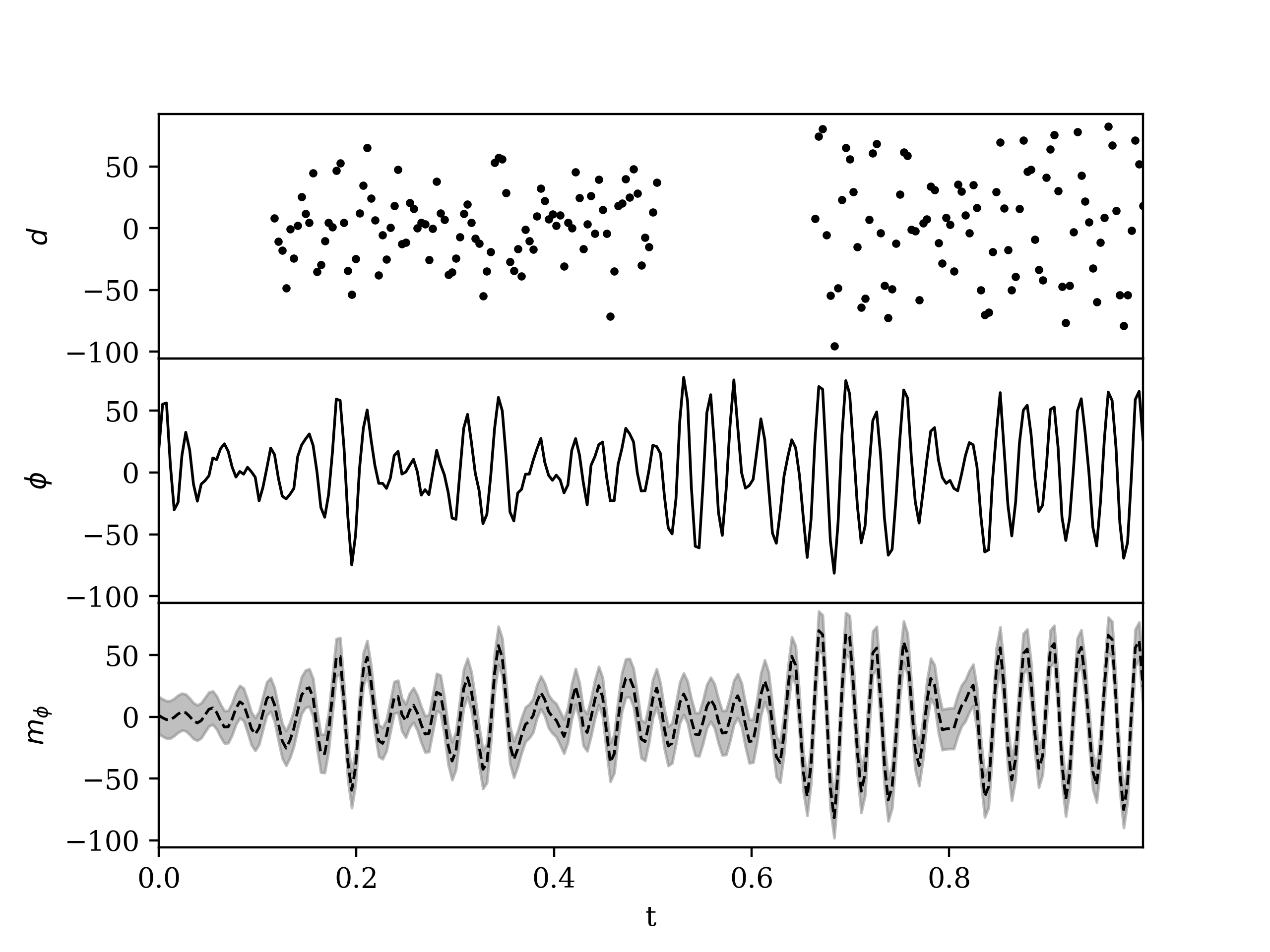}
	\centering
	\caption{Data $d$ (top), signal $\phi$ (middle) and reconstruction $m_\phi$ (bottom) using the MAP estimate of the spectrum. The gray area denotes the one sigma uncertainty of the reconstruction.}\label{fig:phirec}
	
\end{figure}

\begin{figure}[ht]
	\includegraphics[scale=0.5, angle=0]{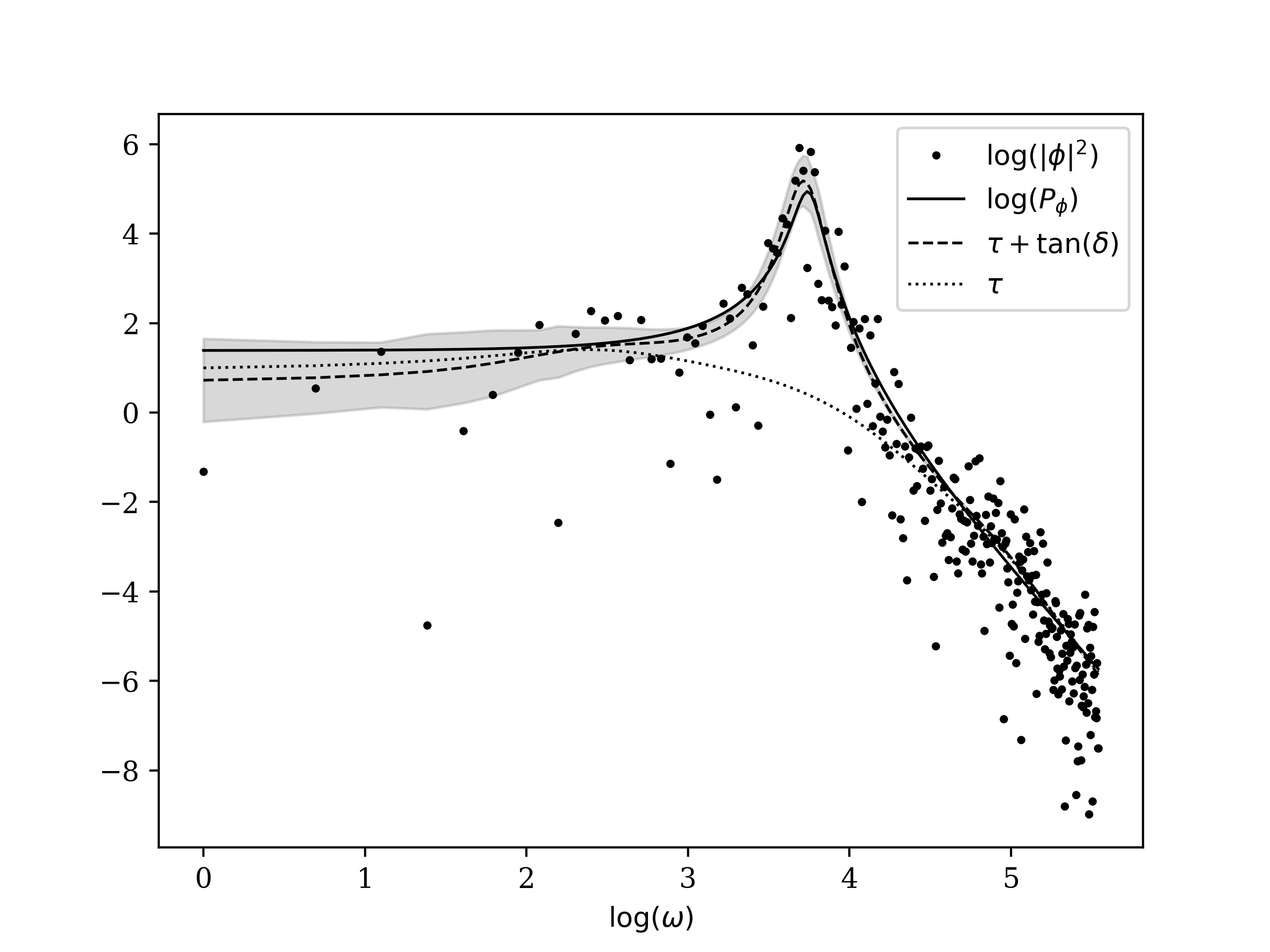}
	
	\centering
	\caption{Reconstruction of $P_\phi$, given $\phi$ (Fig.\,\ref{fig:phirec}, middle panel), on a log-log-scale. The solid line is the theoretical spectrum corresponding to Eq.\,\ref{eq:spec} and the black dots display the spectrum of the actual sample $\phi$. The dashed line is the maximum a posterior estimate obtained by maximizing Eq.\,\ref{eq:h1} and the dotted line is the MAP estimate of $\tau$ alone. The gray area denotes the one sigma uncertainties of the reconstruction. Details on the uncertainty maps are described in appendix \ref{ap:grad}. }\label{fig:recphi}

\end{figure}

\section{Application}\label{sec:application}
For the first consistency check we restrict the analysis to one dimension, the time axis. Consider a differential equation of the form
\begin{equation}\label{eq:osz}
	(\alpha \ \partial_t^2 + \beta \ \partial_t + m^2) \ \phi = \xi  \ ,
\end{equation}
with $(\alpha,\beta,m^2) = (0.0003,0.001,0.5)$. This is the stochastic version of a damped harmonic oscillator. If we assume $\xi$ to be a white noise process with covariance $\Theta=\mathds{1}$, the spectral density of $\phi$ becomes
\begin{equation}\label{eq:spec}
	P_\phi(\omega) = \frac{1}{(\gamma - \alpha \ \omega^2)^2 + (\beta \ \omega)^2} \ .
\end{equation}
A signal $\phi$ (displayed in center panel of Fig.\,\ref{fig:phirec}) can then be generated by drawing one sample from the probability distribution corresponding to $P_\phi$. Assuming that one is given $\phi$, Eq.\,\ref{eq:h1} can be used directly to infer the spectrum by maximizing the corresponding Hamiltonian. In this application the hyper-prior values are set to $(\sigma,\mu,\nu)=(2.0,2.0,0.5 \pi)$. A Gaussian approximation of the uncertainty of the reconstruction is given in terms of the second derivatives of the Hamiltonian. Details are described in appendix \ref{ap:grad}. The results of the reconstruction are shown in Fig.\,\ref{fig:recphi}, where we depict $P_\phi$ as well as the reconstruction on a log-log-scale. One can see that both fields behave as expected, i.e. $\tau$ models the smooth background of the spectrum while $\delta$ reconstructs the divergence and is zero everywhere else. In addition, one sees that the assumption of linearity is true up to the divergent part, as expected.

\begin{figure}[h]
	\includegraphics[scale=0.5, angle=0]{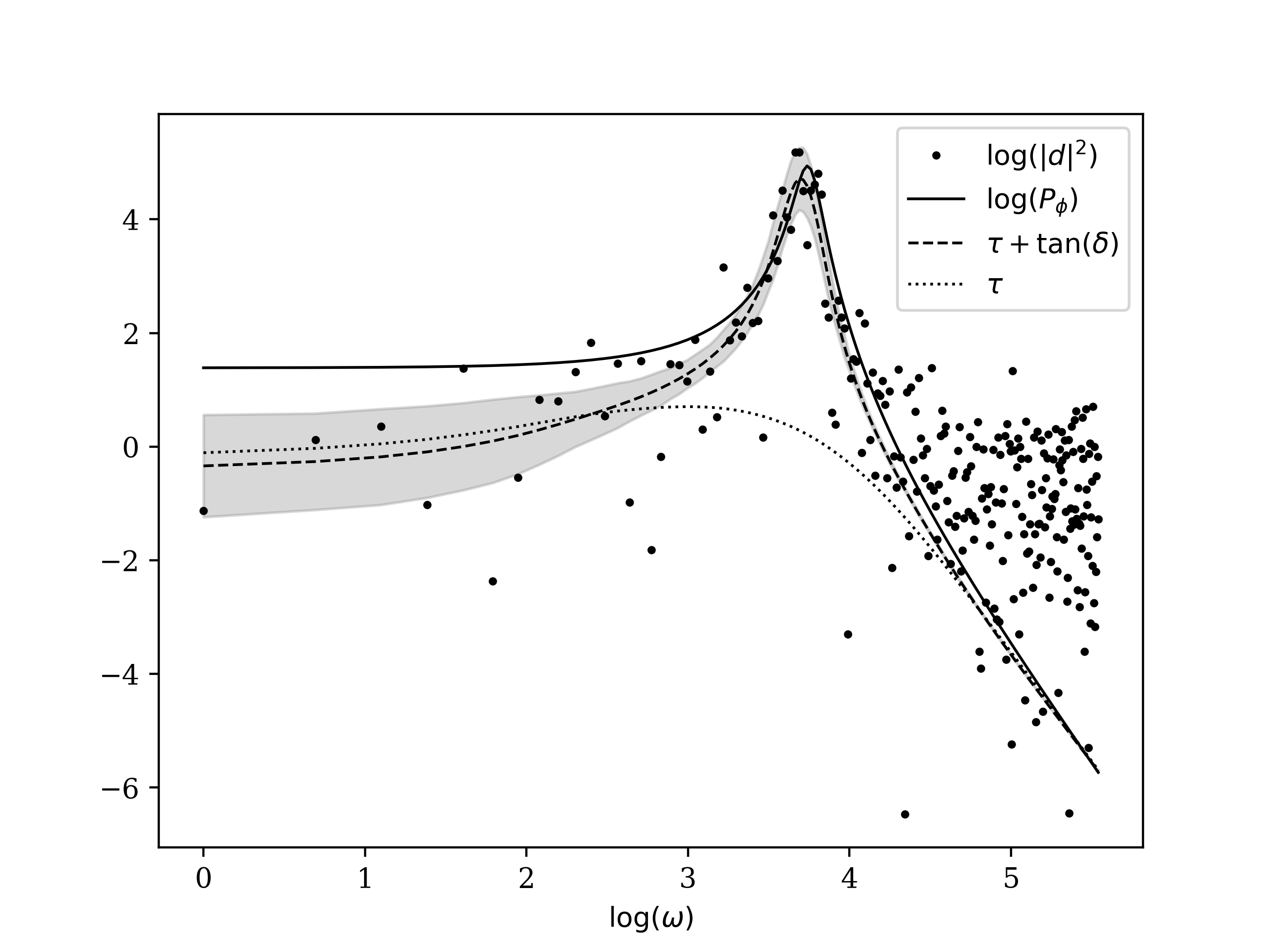}
	
	\centering
	\caption{Reconstruction of $P_\phi$ given noisy data $d$ (Fig.\,\ref{fig:phirec}, top panel). We note that due to the Gaussian approximation the uncertainty estimates are significantly underestimated, in regions of low power.}\label{fig:recda}
	
\end{figure}

Now we assume that instead of $\phi$ we are only given noisy and incomplete measurements $d$ of $\phi$, as shown in Fig.\,\ref{fig:phirec}. We generate mock data assuming Gaussian noise with variance $\sigma_n = 16$. To mimic a more realistic measurement scenario we built the response operator $R$ such that it only measures the signal for certain time intervals. This means that $R$ masks the signal in certain regions and the resulting data, displayed in the top panel of Fig.\,\ref{fig:phirec}, is a noisy and incomplete version of the original signal.

As described in section \ref{sec:post2}, we first reconstruct the spectrum by minimizing the marginal Hamiltonian (Eq.\,\ref{eq:marginalhamilton}) with respect to $\tau$ and $\delta$ to get maximum a posterior solutions. Thereby we used the same values for the hyper-priors as in the perfect data case. The results are displayed in Fig.\,\ref{fig:recda}. Using the results from Eq.\,\ref{eq:empbayes} we also obtain a reconstruction of the original signal $\phi$ as well as corresponding uncertainties (lower panel of Fig.\,\ref{fig:phirec}). We see that even in regions of no data, we partially infer the correct signal since we were able to obtain a good reconstruction of the spectrum in the first place. The quality of these inter- and extrapolations of the mean reconstruction strongly depends on the correlation length of the corresponding dynamical process. This means that if the process is dominated by random excitations rather than deterministic evolution, interpolation over length scales much larger than the correlation length is in principle not possible. However, due to the fact that we infer the full statistics of the process, even in these cases we are able to state the probability of each possible interpolation in terms of the posterior distribution. Due to the empirical Bayes approach spectral uncertainties are ignored for the reconstruction of $\phi$. Therefore, posterior uncertainties of $\phi$ are underestimated, particularly concerning the overall power of the oscillations. This becomes obvious in the regions of no data.

\subsection{Spatio-Temporal Evolution}
\begin{figure*}[htp]
	\includegraphics[scale=0.9, angle=0]{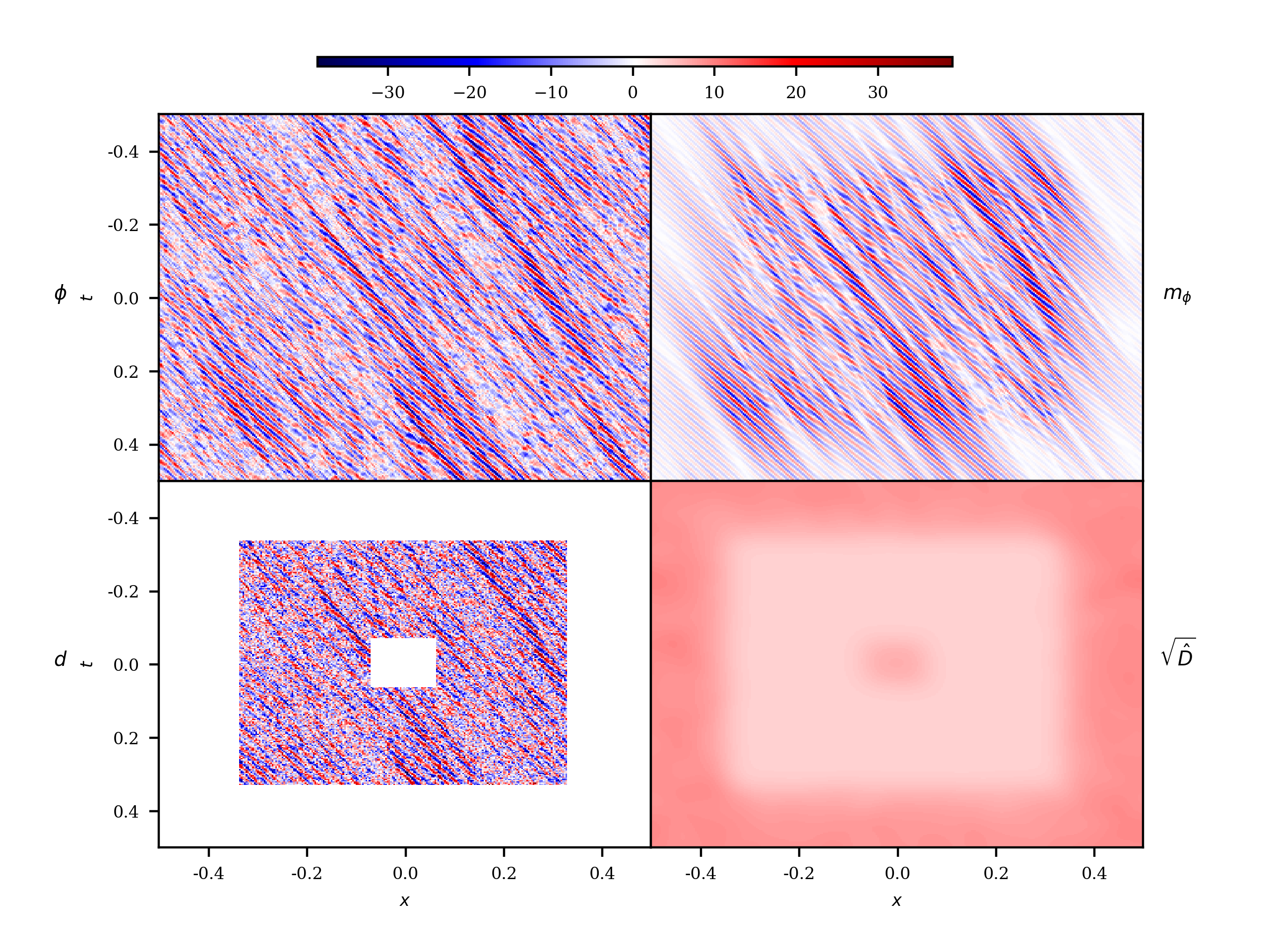}
	
	\centering
	\caption{Signal $\phi$ (top-left, drawn form the process corresponding to Eq.\,\ref{eq:2d}), resulting noisy measurement data $d$ (bottom-left), reconstruction $m_{\phi}$ (top-right) and uncertainty map $\sqrt{\hat{D}}$ (bottom-right) of the reconstruction.}\label{fig:2dphi}
	
	\includegraphics[scale=0.9, angle=0]{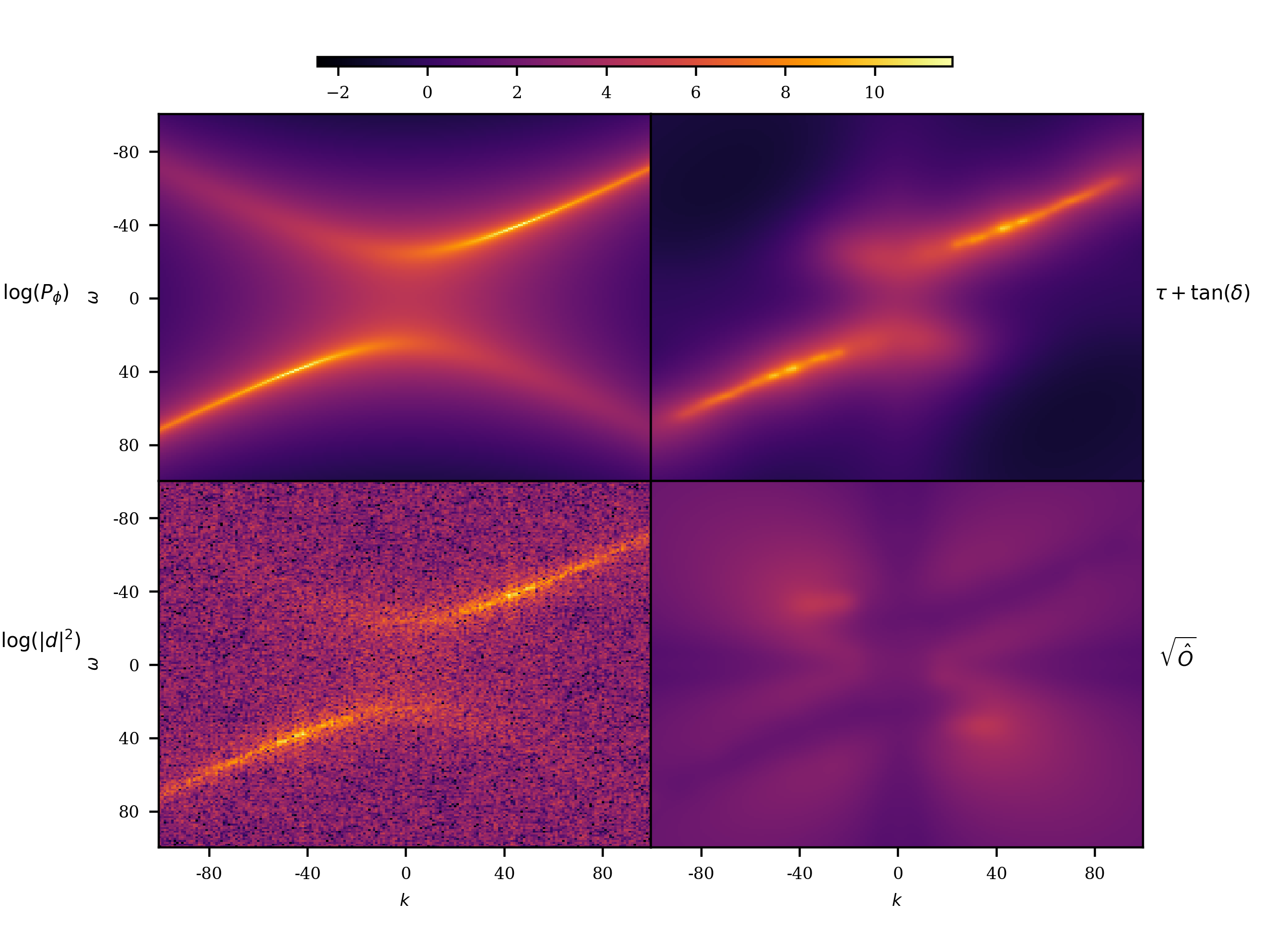}
	\centering
	\caption{For the field and data shown in Fig.\,\ref{fig:2dphi}, logarithmic spectrum $\log(P_\phi)$ (top-left), projected data $\log\left( \left| d\right|^2\right)$ (bottom-left), reconstruction $\tau+\tan(\delta)$ (top-right) and uncertainty estimate $\sqrt{\hat{O}}$ (bottom-right) as defined in appendix \ref{ap:grad}.}\label{fig:2dspec}
\end{figure*}

\begin{figure*}[htp]
	\includegraphics[scale=0.22, angle=0]{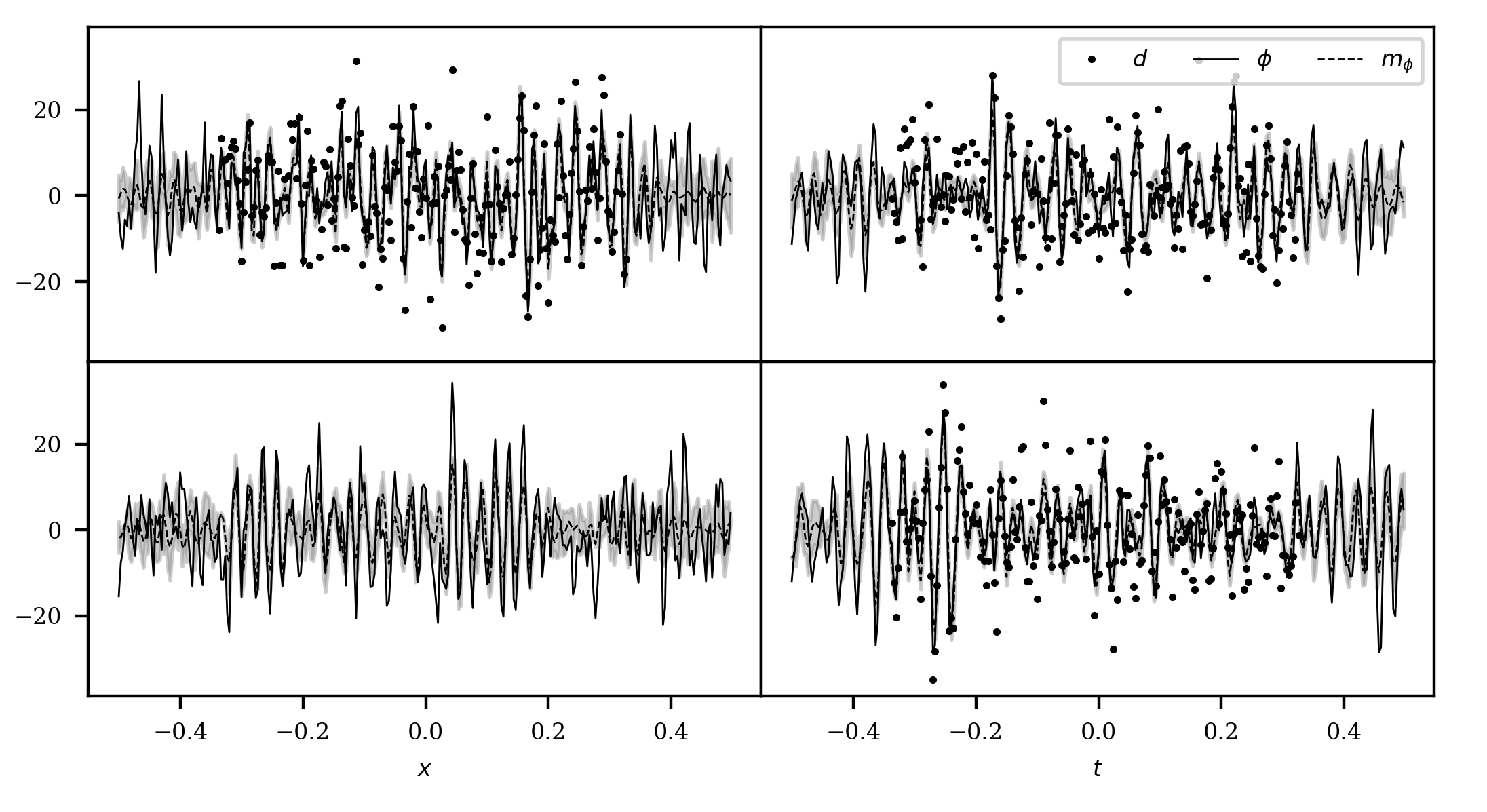}
	
	\centering
	\caption{Slices through the full field, the data and the reconstruction, shown in Fig.\,\ref{fig:2dphi}. The left panels show the spatial structure for $t=-0.27$ (top-left) and $t=0.38$ (bottom-left). The right panels show the temporal evolution at $x=-0.15$ (top-right) and at $x=0.17$ (bottom-right).}\label{fig:cutphi}
	
	\includegraphics[scale=0.22, angle=0]{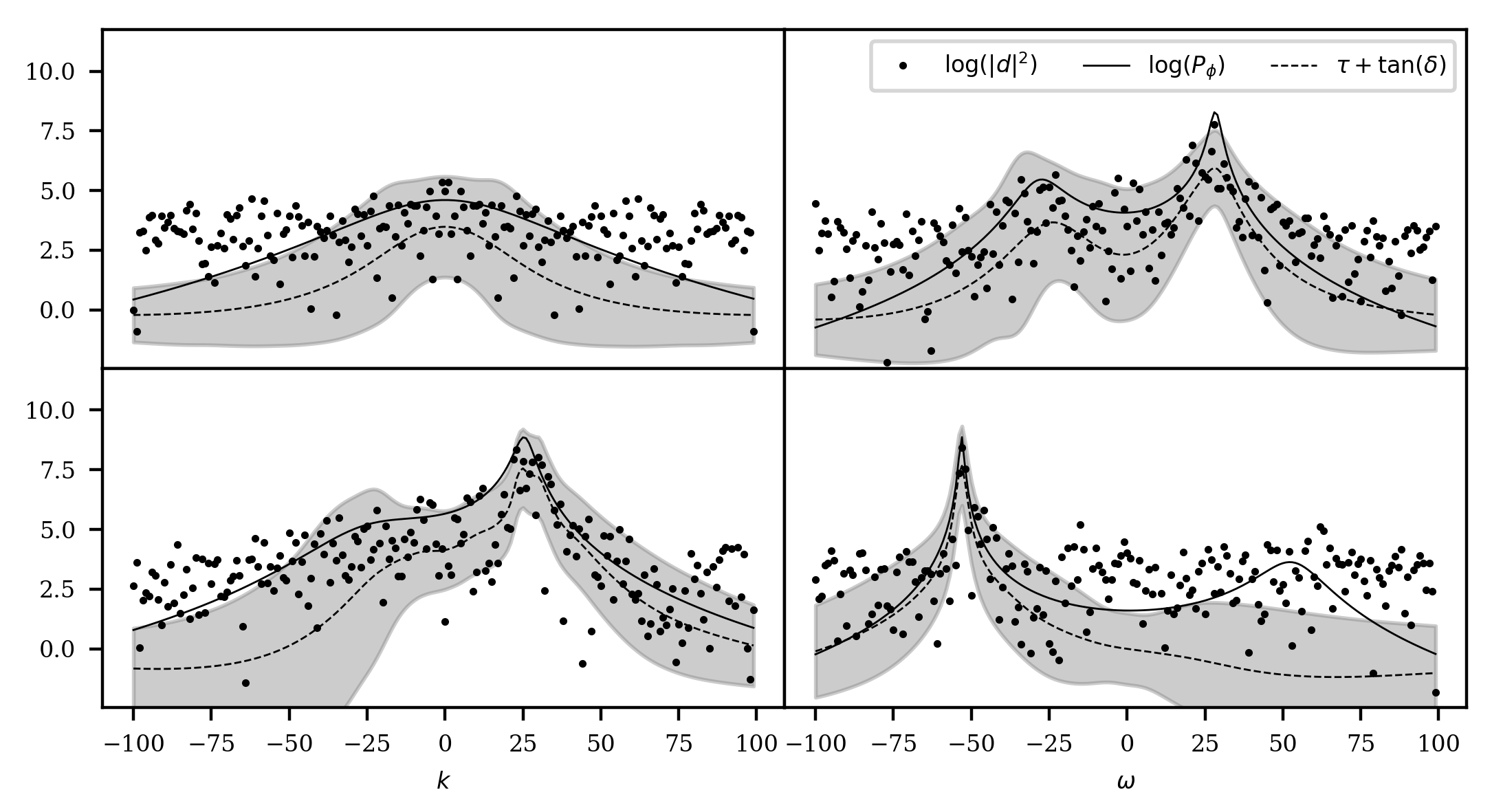}
	
	\centering
	\caption{Slices through the spectrum shown in Fig.\,\ref{fig:2dspec} for fixed $k$ and fixed $\omega$. Top-left: $\omega=0$, bottom-left: $\omega=-30$, top-right $k=-20$ and bottom-right: $k=70$.}\label{fig:cutspec}
\end{figure*}

In the next example we extend the analysis to two dimensions, a spatial and a temporal one. The analysis follows the same spirit as described in the previous section. Using a stochastic process of the form
\begin{equation}\label{eq:2d}
	(\alpha \partial_t^2 - \beta \partial_x^2 - \gamma \partial_x - \rho \partial_t + m^2 ) \phi = \xi \ ,
\end{equation}
we first generate mock data $d$ (bottom-left panel of Fig.\,\ref{fig:2dphi}) from a signal $\phi$  (top-left panel of Fig.\,\ref{fig:2dphi}) using a noise variance of $\sigma_n = 7$. $\xi$ is again a white noise process and $(\alpha, \beta, \gamma, \rho,m^2) = (0.00007,0.0002,0.0014,0.0012,0.1)$. The MAP estimate of the spectrum as well as the spectrum itself is displayed in Fig.\,\ref{fig:2dspec}. In this case, the hyper-priors are set to $(\sigma,\mu,\nu)=(2.5,2.5,0.5 \pi)$. We see that in this setting we are able to recover the dominant features of the spectrum, while features of lower power are not recoverable due to noise. This becomes even clearer when we look at Fig.\,\ref{fig:cutspec}. Here we present slices through the spectrum for different frequency values of $k$ and $\omega$. We see that all features with significant power above the noise level are reconstructed well, while features which are indistinguishable of the noise get suppressed in the reconstruction.

Using the MAP estimate for the spectrum we also reconstruct $\phi$ itself, as displayed in the top-right panel of Fig.\,\ref{fig:2dphi}. We see that even in regions of no data, we reconstruct the dominant oscillations of the system, while small scale structures cannot be recovered. Again, in Fig.\,\ref{fig:cutphi}, we present slices of the data, the signal and the reconstruction for different time-steps and at different locations. The bottom-left subplot shows the spatial structure at a very late time-step, namely in a region where no measurement was made at all. This means that the reconstruction is based entirely on the spectral reconstruction, which serves as a prior, and the constraints which come from data at previous time-steps. Nevertheless, a reasonable estimate of the field configuration is still possible for a certain period after the last measurements.

\begin{figure*}[htp]
	\includegraphics[scale=0.9, angle=0]{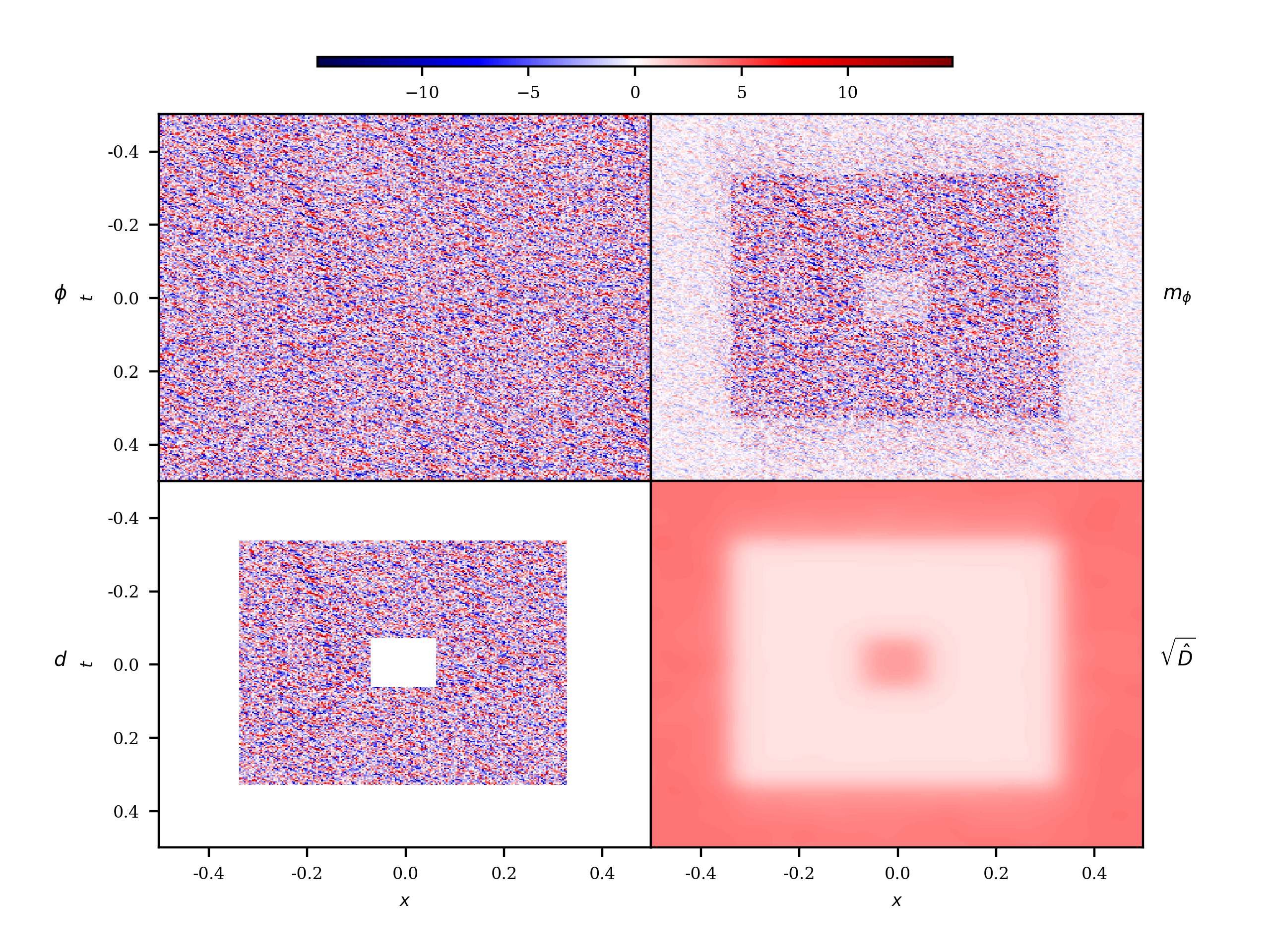}
	
	\centering
	\caption{Signal $\phi$ (top-left), noisy measurement data $d$ (bottom-left), reconstruction $m_{\phi}$ (top-right), and uncertainty map $\sqrt{\hat{D}}$ (bottom-right), for the highly structured spectrum given by Eq.\,\ref{eq:highstr}}\label{fig:2dp}
	
	\includegraphics[scale=0.9, angle=0]{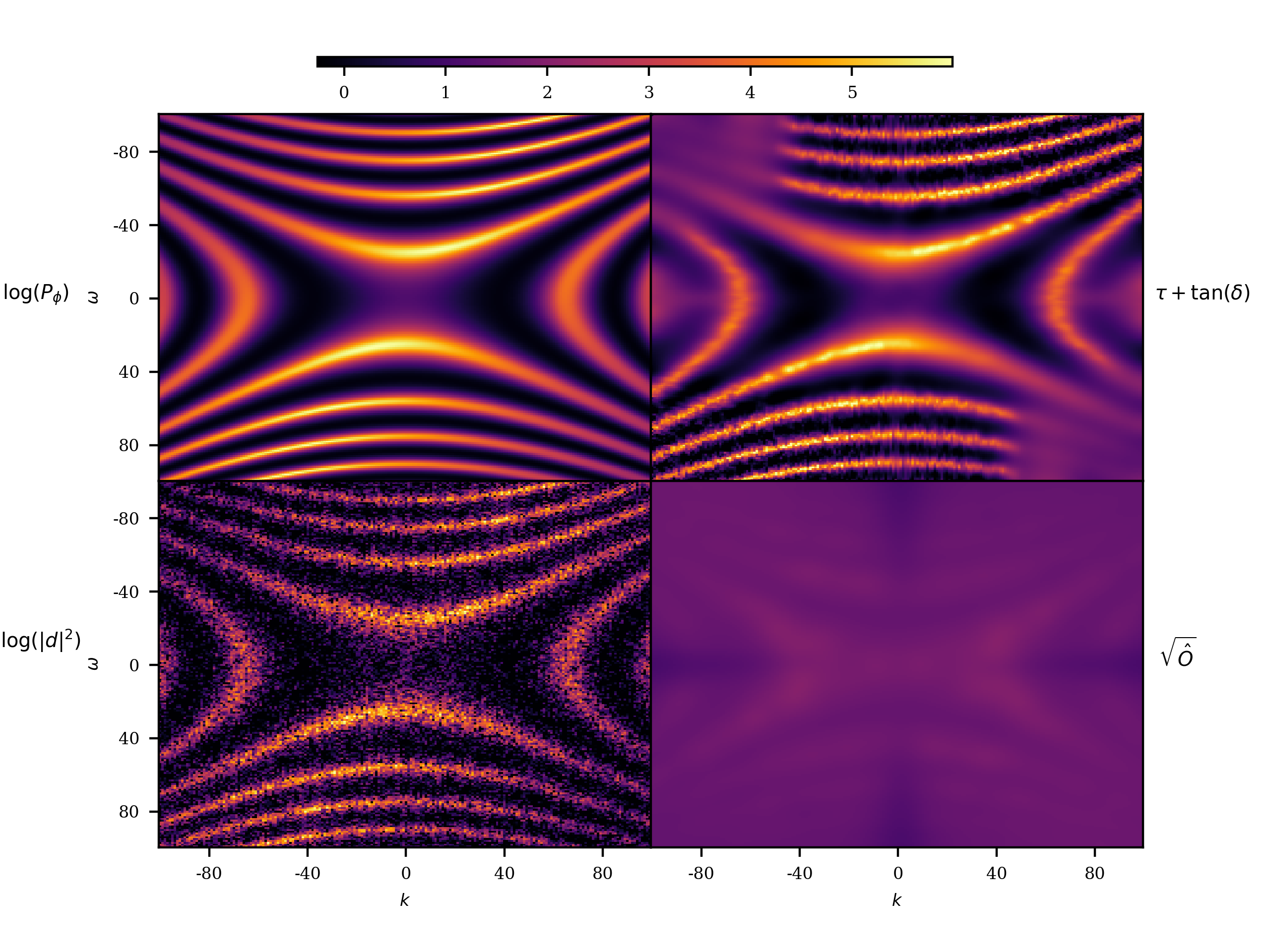}
	
	\centering
	\caption{For the field and data shown in Fig.\,\ref{fig:2dp}, logarithmic spectrum $\log(P_\phi)$ (top-left), projected data $\log\left( \left| d\right|^2\right)$ (bottom-left), reconstruction $\tau+\tan(\delta)$ (top-right) and uncertainty estimate $\sqrt{\hat{O}}$ (bottom-right).}\label{fig:2ds}
\end{figure*}

To demonstrate the advantage of this non-parametric approach we apply the analysis to a highly structured setting, namely with a spectrum of the form
\begin{equation}\label{eq:highstr}
	P_\phi(k,\omega) = \frac{2}{\left( m^2-\sin\left( \alpha k^2 - \beta \omega^2\right)\right)^2 + (\gamma k+\rho \omega)^2 } \ ,
\end{equation}
with $(m^2,\alpha,\beta,\gamma,\rho) = (1.1,0.0025,0.0011,0.002,0.004)$. Although this spectrum is completely artificial, we note that similar periodic and highly-structured spectra also exist in reality. Such are observed, for example, in helioseismology \cite{1988RvMP...60..297B}.
We use the same setup as described in the previous example but reduce the noise variance to $\sigma_n=1$ in order to capture more structure of the spectrum. In addition, the hyper-prior parameters were set to $(\sigma,\mu,\nu)=(4.0,4.0,0.5 \pi)$. The results are shown in Fig.\,\ref{fig:2dp} and Fig.\,\ref{fig:2ds}.

\section{Conclusion}\label{sec:conclusion}
Spectral density estimation is a powerful technique for field dynamics inference.
In this work, posterior distributions for the spectral density and the field itself have been derived which ultimately are used to retrieve maximum a posteriori estimates as well as corresponding uncertainties. The one- and two-dimensional tests indicate that the method behaves as expected in self-consistent scenarios.

Possible applications of this method involve fields which have a non-trivial entanglement between spatial and temporal evolution. One example is the inference of the dynamics of a plasma from observations. Another example is the area of numerical simulations. In particular in astrophysical applications one is often interested in the dynamics of fields that are computationally too expensive to simulate. Treating a few expensive simulations as observations of the field of interest, our method can provide an approximate dynamics, that can mimic essential properties of the real evolution of the field.

The distinction between smooth and divergent parts of the spectrum as well as the corresponding properties of the prior choices for responsible fields $\tau$ and $\delta$ appears to be reasonable in this setting. However, other choices may also be possible. A future goal would be to study other parameterizations of the spectrum in terms of fields in particular in a 4D high resolution setting where a reconstruction of the full spectrum exceeds the range of computability.

Despite the fact that linear autonomous SDEs are an important class of SDEs to study dynamical evolution, an extension to non-autonomous as well as non-linear problems is a desired goal for future work. However, for these cases it appears to be indispensable to have a general method for linear processes first. This is provided by this work.

\begin{acknowledgments}
	We acknowledge helpful discussions and comments
	on the manuscript by Jakob Knollm{\"u}ller, Reimar Leike, Philipp Arras and Martin Dupont.
\end{acknowledgments}

\newpage

\bibliographystyle{apsrev4-1}
\bibliography{SDE}

\begin{thebibliography}{24}%
\makeatletter
\providecommand \@ifxundefined [1]{%
 \@ifx{#1\undefined}
}%
\providecommand \@ifnum [1]{%
 \ifnum #1\expandafter \@firstoftwo
 \else \expandafter \@secondoftwo
 \fi
}%
\providecommand \@ifx [1]{%
 \ifx #1\expandafter \@firstoftwo
 \else \expandafter \@secondoftwo
 \fi
}%
\providecommand \natexlab [1]{#1}%
\providecommand \enquote  [1]{``#1''}%
\providecommand \bibnamefont  [1]{#1}%
\providecommand \bibfnamefont [1]{#1}%
\providecommand \citenamefont [1]{#1}%
\providecommand \href@noop [0]{\@secondoftwo}%
\providecommand \href [0]{\begingroup \@sanitize@url \@href}%
\providecommand \@href[1]{\@@startlink{#1}\@@href}%
\providecommand \@@href[1]{\endgroup#1\@@endlink}%
\providecommand \@sanitize@url [0]{\catcode `\\12\catcode `\$12\catcode
  `\&12\catcode `\#12\catcode `\^12\catcode `\_12\catcode `\%12\relax}%
\providecommand \@@startlink[1]{}%
\providecommand \@@endlink[0]{}%
\providecommand \url  [0]{\begingroup\@sanitize@url \@url }%
\providecommand \@url [1]{\endgroup\@href {#1}{\urlprefix }}%
\providecommand \urlprefix  [0]{URL }%
\providecommand \Eprint [0]{\href }%
\providecommand \doibase [0]{http://dx.doi.org/}%
\providecommand \selectlanguage [0]{\@gobble}%
\providecommand \bibinfo  [0]{\@secondoftwo}%
\providecommand \bibfield  [0]{\@secondoftwo}%
\providecommand \translation [1]{[#1]}%
\providecommand \BibitemOpen [0]{}%
\providecommand \bibitemStop [0]{}%
\providecommand \bibitemNoStop [0]{.\EOS\space}%
\providecommand \EOS [0]{\spacefactor3000\relax}%
\providecommand \BibitemShut  [1]{\csname bibitem#1\endcsname}%
\let\auto@bib@innerbib\@empty
\bibitem [{\citenamefont {Henderson}\ and\ \citenamefont
  {Plaschko}(2006)}]{henderson2006stochastic}%
  \BibitemOpen
  \bibfield  {author} {\bibinfo {author} {\bibfnamefont {D.}~\bibnamefont
  {Henderson}}\ and\ \bibinfo {author} {\bibfnamefont {P.}~\bibnamefont
  {Plaschko}},\ }\href {https://books.google.de/books?id=lqYoAQAAMAAJ} {\emph
  {\bibinfo {title} {Stochastic Differential Equations in Science and
  Engineering}}},\ \bibinfo {series} {Stochastic Differential Equations in
  Science and Engineering}\ No.\ \bibinfo {number} {Bd. 1}\ (\bibinfo
  {publisher} {World Scientific Pub.},\ \bibinfo {year} {2006})\BibitemShut
  {NoStop}%
\bibitem [{\citenamefont {Iacus}(2008)}]{Iacus:2008:SIS:1386239}%
  \BibitemOpen
  \bibfield  {author} {\bibinfo {author} {\bibfnamefont {S.~M.}\ \bibnamefont
  {Iacus}},\ }\href@noop {} {\emph {\bibinfo {title} {Simulation and Inference
  for Stochastic Differential Equations: With R Examples (Springer Series in
  Statistics)}}},\ \bibinfo {edition} {1st}\ ed.\ (\bibinfo  {publisher}
  {Springer Publishing Company, Incorporated},\ \bibinfo {year}
  {2008})\BibitemShut {NoStop}%
\bibitem [{\citenamefont {{Klus}}\ \emph {et~al.}(2017)\citenamefont {{Klus}},
  \citenamefont {{N{\"u}ske}}, \citenamefont {{Koltai}}, \citenamefont {{Wu}},
  \citenamefont {{Kevrekidis}}, \citenamefont {{Sch{\"u}tte}},\ and\
  \citenamefont {{No{\'e}}}}]{2017arXiv170310112K}%
  \BibitemOpen
  \bibfield  {author} {\bibinfo {author} {\bibfnamefont {S.}~\bibnamefont
  {{Klus}}}, \bibinfo {author} {\bibfnamefont {F.}~\bibnamefont {{N{\"u}ske}}},
  \bibinfo {author} {\bibfnamefont {P.}~\bibnamefont {{Koltai}}}, \bibinfo
  {author} {\bibfnamefont {H.}~\bibnamefont {{Wu}}}, \bibinfo {author}
  {\bibfnamefont {I.}~\bibnamefont {{Kevrekidis}}}, \bibinfo {author}
  {\bibfnamefont {C.}~\bibnamefont {{Sch{\"u}tte}}}, \ and\ \bibinfo {author}
  {\bibfnamefont {F.}~\bibnamefont {{No{\'e}}}},\ }\href@noop {} {\bibfield
  {journal} {\bibinfo  {journal} {ArXiv e-prints}\ } (\bibinfo {year}
  {2017})},\ \Eprint {http://arxiv.org/abs/1703.10112} {arXiv:1703.10112
  [math.DS]} \BibitemShut {NoStop}%
\bibitem [{\citenamefont {Roweis}\ and\ \citenamefont
  {Ghahramani}(1999)}]{Roweis:1999:URL:309394.309396}%
  \BibitemOpen
  \bibfield  {author} {\bibinfo {author} {\bibfnamefont {S.}~\bibnamefont
  {Roweis}}\ and\ \bibinfo {author} {\bibfnamefont {Z.}~\bibnamefont
  {Ghahramani}},\ }\href {\doibase 10.1162/089976699300016674} {\bibfield
  {journal} {\bibinfo  {journal} {Neural Comput.}\ }\textbf {\bibinfo {volume}
  {11}},\ \bibinfo {pages} {305} (\bibinfo {year} {1999})}\BibitemShut
  {NoStop}%
\bibitem [{\citenamefont {Potiron}\ and\ \citenamefont
  {Mykland}(2016)}]{RePEc:arx:papers:1603.05700}%
  \BibitemOpen
  \bibfield  {author} {\bibinfo {author} {\bibfnamefont {Y.}~\bibnamefont
  {Potiron}}\ and\ \bibinfo {author} {\bibfnamefont {P.}~\bibnamefont
  {Mykland}},\ }\href {https://ideas.repec.org/p/arx/papers/1603.05700.html}
  {\emph {\bibinfo {title} {{Local Parametric Estimation in High Frequency
  Data}}}},\ \bibinfo {type} {Papers}\ \bibinfo {number} {1603.05700}\
  (\bibinfo  {institution} {arXiv.org},\ \bibinfo {year} {2016})\BibitemShut
  {NoStop}%
\bibitem [{\citenamefont {Spangl}\ and\ \citenamefont
  {Dutter}(2013)}]{Spangl2013}%
  \BibitemOpen
  \bibfield  {author} {\bibinfo {author} {\bibfnamefont {B.}~\bibnamefont
  {Spangl}}\ and\ \bibinfo {author} {\bibfnamefont {R.}~\bibnamefont
  {Dutter}},\ }\enquote {\bibinfo {title} {Robustness in time series: Robust
  frequency domain analysis},}\ in\ \href {\doibase
  10.1007/978-3-642-35494-6_13} {\emph {\bibinfo {booktitle} {Robustness and
  Complex Data Structures: Festschrift in Honour of Ursula Gather}}},\ \bibinfo
  {editor} {edited by\ \bibinfo {editor} {\bibfnamefont {C.}~\bibnamefont
  {Becker}}, \bibinfo {editor} {\bibfnamefont {R.}~\bibnamefont {Fried}}, \
  and\ \bibinfo {editor} {\bibfnamefont {S.}~\bibnamefont {Kuhnt}}}\ (\bibinfo
  {publisher} {Springer Berlin Heidelberg},\ \bibinfo {address} {Berlin,
  Heidelberg},\ \bibinfo {year} {2013})\ pp.\ \bibinfo {pages}
  {207--223}\BibitemShut {NoStop}%
\bibitem [{\citenamefont {Ruttor}\ \emph {et~al.}(2013)\citenamefont {Ruttor},
  \citenamefont {Batz},\ and\ \citenamefont
  {Opper}}]{DBLP:conf/nips/RuttorBO13}%
  \BibitemOpen
  \bibfield  {author} {\bibinfo {author} {\bibfnamefont {A.}~\bibnamefont
  {Ruttor}}, \bibinfo {author} {\bibfnamefont {P.}~\bibnamefont {Batz}}, \ and\
  \bibinfo {author} {\bibfnamefont {M.}~\bibnamefont {Opper}},\ }in\ \href
  {http://papers.nips.cc/paper/4967-approximate-gaussian-process-inference-for-the-drift-function-in-stochastic-differential-equations}
  {\emph {\bibinfo {booktitle} {Advances in Neural Information Processing
  Systems 26: 27th Annual Conference on Neural Information Processing Systems
  2013. Proceedings of a meeting held December 5-8, 2013, Lake Tahoe, Nevada,
  United States.}}}\ (\bibinfo {year} {2013})\ pp.\ \bibinfo {pages}
  {2040--2048}\BibitemShut {NoStop}%
\bibitem [{\citenamefont {Bretthorst}(1988)}]{bretthorst1988bayesian}%
  \BibitemOpen
  \bibfield  {author} {\bibinfo {author} {\bibfnamefont {G.}~\bibnamefont
  {Bretthorst}},\ }\href {https://books.google.de/books?id=fPm8OU6li00C} {\emph
  {\bibinfo {title} {Bayesian spectrum analysis and parameter estimation}}},\
  Lecture notes in statistics\ (\bibinfo  {publisher} {Springer-Verlag},\
  \bibinfo {year} {1988})\BibitemShut {NoStop}%
\bibitem [{\citenamefont {Macaro}\ and\ \citenamefont
  {Prado}(2014)}]{Macaro2014}%
  \BibitemOpen
  \bibfield  {author} {\bibinfo {author} {\bibfnamefont {C.}~\bibnamefont
  {Macaro}}\ and\ \bibinfo {author} {\bibfnamefont {R.}~\bibnamefont {Prado}},\
  }\href {\doibase 10.1007/s11336-013-9354-0} {\bibfield  {journal} {\bibinfo
  {journal} {Psychometrika}\ }\textbf {\bibinfo {volume} {79}},\ \bibinfo
  {pages} {105} (\bibinfo {year} {2014})}\BibitemShut {NoStop}%
\bibitem [{\citenamefont {Knauf}\ \emph {et~al.}(2016)\citenamefont {Knauf},
  \citenamefont {Memmert},\ and\ \citenamefont {Brefeld}}]{Knauf2016}%
  \BibitemOpen
  \bibfield  {author} {\bibinfo {author} {\bibfnamefont {K.}~\bibnamefont
  {Knauf}}, \bibinfo {author} {\bibfnamefont {D.}~\bibnamefont {Memmert}}, \
  and\ \bibinfo {author} {\bibfnamefont {U.}~\bibnamefont {Brefeld}},\ }\href
  {\doibase 10.1007/s10994-015-5520-1} {\bibfield  {journal} {\bibinfo
  {journal} {Machine Learning}\ }\textbf {\bibinfo {volume} {102}},\ \bibinfo
  {pages} {247} (\bibinfo {year} {2016})}\BibitemShut {NoStop}%
\bibitem [{\citenamefont {Subba~Rao}\ and\ \citenamefont
  {Terdik}()}]{JTSA:JTSA12245}%
  \BibitemOpen
  \bibfield  {author} {\bibinfo {author} {\bibfnamefont {T.}~\bibnamefont
  {Subba~Rao}}\ and\ \bibinfo {author} {\bibfnamefont {G.}~\bibnamefont
  {Terdik}},\ }\href {\doibase 10.1111/jtsa.12245} {\bibfield  {journal}
  {\bibinfo  {journal} {Journal of Time Series Analysis}\ ,\ \bibinfo {pages}
  {n/a}}}\bibinfo {note} {10.1111/jtsa.12245}\BibitemShut {NoStop}%
\bibitem [{\citenamefont {Zheng}\ \emph {et~al.}(2010)\citenamefont {Zheng},
  \citenamefont {Zhu},\ and\ \citenamefont {Roy}}]{doi:10.1093/biomet/asp066}%
  \BibitemOpen
  \bibfield  {author} {\bibinfo {author} {\bibfnamefont {Y.}~\bibnamefont
  {Zheng}}, \bibinfo {author} {\bibfnamefont {J.}~\bibnamefont {Zhu}}, \ and\
  \bibinfo {author} {\bibfnamefont {A.}~\bibnamefont {Roy}},\ }\href {\doibase
  10.1093/biomet/asp066} {\bibfield  {journal} {\bibinfo  {journal}
  {Biometrika}\ }\textbf {\bibinfo {volume} {97}},\ \bibinfo {pages} {238}
  (\bibinfo {year} {2010})}\BibitemShut {NoStop}%
\bibitem [{\citenamefont
  {{En{\ss}lin}}(2013{\natexlab{a}})}]{2013AIPC.1553..184E}%
  \BibitemOpen
  \bibfield  {author} {\bibinfo {author} {\bibfnamefont {T.}~\bibnamefont
  {{En{\ss}lin}}},\ }in\ \href {\doibase 10.1063/1.4819999} {\emph {\bibinfo
  {booktitle} {American Institute of Physics Conference Series}}},\ \bibinfo
  {series} {American Institute of Physics Conference Series}, Vol.\ \bibinfo
  {volume} {1553},\ \bibinfo {editor} {edited by\ \bibinfo {editor}
  {\bibfnamefont {U.}~\bibnamefont {{von Toussaint}}}}\ (\bibinfo {year}
  {2013})\ pp.\ \bibinfo {pages} {184--191},\ \Eprint
  {http://arxiv.org/abs/1301.2556} {arXiv:1301.2556 [astro-ph.IM]} \BibitemShut
  {NoStop}%
\bibitem [{\citenamefont {{En{\ss}lin}}\ \emph {et~al.}(2009)\citenamefont
  {{En{\ss}lin}}, \citenamefont {{Frommert}},\ and\ \citenamefont
  {{Kitaura}}}]{2009PhRvD..80j5005E}%
  \BibitemOpen
  \bibfield  {author} {\bibinfo {author} {\bibfnamefont {T.~A.}\ \bibnamefont
  {{En{\ss}lin}}}, \bibinfo {author} {\bibfnamefont {M.}~\bibnamefont
  {{Frommert}}}, \ and\ \bibinfo {author} {\bibfnamefont {F.~S.}\ \bibnamefont
  {{Kitaura}}},\ }\href {\doibase 10.1103/PhysRevD.80.105005} {\bibfield
  {journal} {\bibinfo  {journal} {\prd}\ }\textbf {\bibinfo {volume} {80}},\
  \bibinfo {eid} {105005} (\bibinfo {year} {2009})},\ \Eprint
  {http://arxiv.org/abs/0806.3474} {arXiv:0806.3474} \BibitemShut {NoStop}%
\bibitem [{\citenamefont {{Steininger}}\ \emph {et~al.}(2017)\citenamefont
  {{Steininger}}, \citenamefont {{Dixit}}, \citenamefont {{Frank}},
  \citenamefont {{Greiner}}, \citenamefont {{Hutschenreuter}}, \citenamefont
  {{Knollm{\"u}ller}}, \citenamefont {{Leike}}, \citenamefont {{Porqueres}},
  \citenamefont {{Pumpe}}, \citenamefont {{Reinecke}}, \citenamefont {{{\v
  S}raml}}, \citenamefont {{Varady}},\ and\ \citenamefont
  {{En{\ss}lin}}}]{2017arXiv170801073S}%
  \BibitemOpen
  \bibfield  {author} {\bibinfo {author} {\bibfnamefont {T.}~\bibnamefont
  {{Steininger}}}, \bibinfo {author} {\bibfnamefont {J.}~\bibnamefont
  {{Dixit}}}, \bibinfo {author} {\bibfnamefont {P.}~\bibnamefont {{Frank}}},
  \bibinfo {author} {\bibfnamefont {M.}~\bibnamefont {{Greiner}}}, \bibinfo
  {author} {\bibfnamefont {S.}~\bibnamefont {{Hutschenreuter}}}, \bibinfo
  {author} {\bibfnamefont {J.}~\bibnamefont {{Knollm{\"u}ller}}}, \bibinfo
  {author} {\bibfnamefont {R.}~\bibnamefont {{Leike}}}, \bibinfo {author}
  {\bibfnamefont {N.}~\bibnamefont {{Porqueres}}}, \bibinfo {author}
  {\bibfnamefont {D.}~\bibnamefont {{Pumpe}}}, \bibinfo {author} {\bibfnamefont
  {M.}~\bibnamefont {{Reinecke}}}, \bibinfo {author} {\bibfnamefont
  {M.}~\bibnamefont {{{\v S}raml}}}, \bibinfo {author} {\bibfnamefont
  {C.}~\bibnamefont {{Varady}}}, \ and\ \bibinfo {author} {\bibfnamefont
  {T.}~\bibnamefont {{En{\ss}lin}}},\ }\href@noop {} {\bibfield  {journal}
  {\bibinfo  {journal} {ArXiv e-prints}\ } (\bibinfo {year} {2017})},\ \Eprint
  {http://arxiv.org/abs/1708.01073} {arXiv:1708.01073 [astro-ph.IM]}
  \BibitemShut {NoStop}%
\bibitem [{\citenamefont {{En{\ss}lin}}(2014)}]{2014AIPC.1636...49E}%
  \BibitemOpen
  \bibfield  {author} {\bibinfo {author} {\bibfnamefont {T.}~\bibnamefont
  {{En{\ss}lin}}},\ }\href {\doibase 10.1063/1.4903709} {\bibfield  {journal}
  {\bibinfo  {journal} {Bayesian Inference and Maximum Entropy Methods in
  Science and Engineering}\ }\textbf {\bibinfo {volume} {1636}},\ \bibinfo
  {pages} {49} (\bibinfo {year} {2014})},\ \Eprint
  {http://arxiv.org/abs/1405.7701} {arXiv:1405.7701 [astro-ph.IM]} \BibitemShut
  {NoStop}%
\bibitem [{\citenamefont {{Knollm{\"u}ller}}\ and\ \citenamefont
  {{En{\ss}lin}}(2017)}]{2017arXiv170502344K}%
  \BibitemOpen
  \bibfield  {author} {\bibinfo {author} {\bibfnamefont {J.}~\bibnamefont
  {{Knollm{\"u}ller}}}\ and\ \bibinfo {author} {\bibfnamefont {T.~A.}\
  \bibnamefont {{En{\ss}lin}}},\ }\href@noop {} {\bibfield  {journal} {\bibinfo
   {journal} {ArXiv e-prints}\ } (\bibinfo {year} {2017})},\ \Eprint
  {http://arxiv.org/abs/1705.02344} {arXiv:1705.02344 [stat.ME]} \BibitemShut
  {NoStop}%
\bibitem [{\citenamefont
  {{En{\ss}lin}}(2013{\natexlab{b}})}]{2013PhRvE..87a3308E}%
  \BibitemOpen
  \bibfield  {author} {\bibinfo {author} {\bibfnamefont {T.~A.}\ \bibnamefont
  {{En{\ss}lin}}},\ }\href {\doibase 10.1103/PhysRevE.87.013308} {\bibfield
  {journal} {\bibinfo  {journal} {\pre}\ }\textbf {\bibinfo {volume} {87}},\
  \bibinfo {eid} {013308} (\bibinfo {year} {2013}{\natexlab{b}})},\ \Eprint
  {http://arxiv.org/abs/1206.4229} {arXiv:1206.4229 [physics.comp-ph]}
  \BibitemShut {NoStop}%
\bibitem [{\citenamefont {{Pumpe}}\ \emph {et~al.}(2016)\citenamefont
  {{Pumpe}}, \citenamefont {{Greiner}}, \citenamefont {{M{\"u}ller}},\ and\
  \citenamefont {{En{\ss}lin}}}]{2016PhRvE..94a2132P}%
  \BibitemOpen
  \bibfield  {author} {\bibinfo {author} {\bibfnamefont {D.}~\bibnamefont
  {{Pumpe}}}, \bibinfo {author} {\bibfnamefont {M.}~\bibnamefont {{Greiner}}},
  \bibinfo {author} {\bibfnamefont {E.}~\bibnamefont {{M{\"u}ller}}}, \ and\
  \bibinfo {author} {\bibfnamefont {T.~A.}\ \bibnamefont {{En{\ss}lin}}},\
  }\href {\doibase 10.1103/PhysRevE.94.012132} {\bibfield  {journal} {\bibinfo
  {journal} {\pre}\ }\textbf {\bibinfo {volume} {94}},\ \bibinfo {eid} {012132}
  (\bibinfo {year} {2016})},\ \Eprint {http://arxiv.org/abs/1601.07901}
  {arXiv:1601.07901 [physics.data-an]} \BibitemShut {NoStop}%
\bibitem [{\citenamefont {Wiener}()}]{wiener_extrapolation_1950}%
  \BibitemOpen
  \bibfield  {author} {\bibinfo {author} {\bibfnamefont {N.}~\bibnamefont
  {Wiener}},\ }\href {//catalog.hathitrust.org/Record/010056247} {\emph
  {\bibinfo {title} {Extrapolation, interpolation, and smoothing of stationary
  time series, with engineering applications.}}},\ \bibinfo {series}
  {Stationary time series}\ No.\ \bibinfo {number} {ix, 163 p.}\ (\bibinfo
  {publisher} {Technology Press of the Massachusetts Institute
  {ofTechnology}})\BibitemShut {NoStop}%
\bibitem [{\citenamefont {{Oppermann}}\ \emph {et~al.}(2013)\citenamefont
  {{Oppermann}}, \citenamefont {{Selig}}, \citenamefont {{Bell}},\ and\
  \citenamefont {{En{\ss}lin}}}]{2013PhRvE..87c2136O}%
  \BibitemOpen
  \bibfield  {author} {\bibinfo {author} {\bibfnamefont {N.}~\bibnamefont
  {{Oppermann}}}, \bibinfo {author} {\bibfnamefont {M.}~\bibnamefont
  {{Selig}}}, \bibinfo {author} {\bibfnamefont {M.~R.}\ \bibnamefont {{Bell}}},
  \ and\ \bibinfo {author} {\bibfnamefont {T.~A.}\ \bibnamefont
  {{En{\ss}lin}}},\ }\href {\doibase 10.1103/PhysRevE.87.032136} {\bibfield
  {journal} {\bibinfo  {journal} {\pre}\ }\textbf {\bibinfo {volume} {87}},\
  \bibinfo {eid} {032136} (\bibinfo {year} {2013})},\ \Eprint
  {http://arxiv.org/abs/1210.6866} {arXiv:1210.6866 [astro-ph.IM]} \BibitemShut
  {NoStop}%
\bibitem [{\citenamefont {Bishop}(1995)}]{Bishop:1995:NNP:525960}%
  \BibitemOpen
  \bibfield  {author} {\bibinfo {author} {\bibfnamefont {C.~M.}\ \bibnamefont
  {Bishop}},\ }\href@noop {} {\emph {\bibinfo {title} {Neural Networks for
  Pattern Recognition}}}\ (\bibinfo  {publisher} {Oxford University Press,
  Inc.},\ \bibinfo {address} {New York, NY, USA},\ \bibinfo {year}
  {1995})\BibitemShut {NoStop}%
\bibitem [{\citenamefont {{Bahcall}}\ and\ \citenamefont
  {{Ulrich}}(1988)}]{1988RvMP...60..297B}%
  \BibitemOpen
  \bibfield  {author} {\bibinfo {author} {\bibfnamefont {J.~N.}\ \bibnamefont
  {{Bahcall}}}\ and\ \bibinfo {author} {\bibfnamefont {R.~K.}\ \bibnamefont
  {{Ulrich}}},\ }\href {\doibase 10.1103/RevModPhys.60.297} {\bibfield
  {journal} {\bibinfo  {journal} {Reviews of Modern Physics}\ }\textbf
  {\bibinfo {volume} {60}},\ \bibinfo {pages} {297} (\bibinfo {year}
  {1988})}\BibitemShut {NoStop}%
\bibitem [{\citenamefont {Forsythe}\ and\ \citenamefont
  {Wasow}(1960)}]{forsythe1960finite}%
  \BibitemOpen
  \bibfield  {author} {\bibinfo {author} {\bibfnamefont {G.}~\bibnamefont
  {Forsythe}}\ and\ \bibinfo {author} {\bibfnamefont {W.}~\bibnamefont
  {Wasow}},\ }\href {https://books.google.de/books?id=yK0NAQAAIAAJ} {\emph
  {\bibinfo {title} {Finite-difference methods for partial differential
  equations}}},\ Applied mathematics series\ (\bibinfo  {publisher} {Wiley},\
  \bibinfo {year} {1960})\BibitemShut {NoStop}%
\end{thebibliography}%

\begin{appendix}
\section{Smoothness Prior in higher dimensions}\label{ap:smooth}
A smoothness prior in higher dimensions is sometimes constructed by imposing the constraint that, at each point $g$, the log-Laplacian $\partial^2/\partial(\log(g))^2$ of the field should be small. This however, is not sufficient to impose smoothness in cases where one encounters concave and convex curvature along different directions simultaneously (e.g., saddle surfaces). In the following we briefly outline why this is the case.

If we consider the logarithmic Hessian of a field $\psi$ at each point $\mathbf{y} \in \mathds{R}^N$, defined as
\begin{equation}
	(H[\psi])_{ij}(\mathbf{y})= \frac{\partial^2 \psi(\mathbf{y})}{\partial\log(y_i) \ \partial\log(y_j)} \ , \ i,j \in \{1,... ,N\} \ ,
\end{equation}
we notice that the log-Laplacian is equal to the trace of the Hessian and therefore to the sum of the corresponding eigenvalues. This indicates that if the eigenvalues are positive and negative (corresponding to convex and concave curvature along the eigendirections), the Laplacian can become zero even though the surface has nonzero curvature. As a result, saddle-surfaces with equal absolute curvature and constant surfaces are equally likely in the corresponding prior. This is not a desired behavior. The goal of a generic smoothness-prior should be to assign lower probabilities also to surfaces with altering curvature.

We therefore propose to use a prior which aims to minimize all quadratic, logarithmic variations of a field $\psi$ simultaneously. For a M-dimensional space the exponential factor of the prior reads
\begin{equation}
\psi^\dagger T_\sigma^{-1} \psi = \frac{1}{\sigma^2} \ \psi^\dagger   \sum\limits_{i=1}^{M} \left(  T_{ii}^{-1}  + 2 \sum\limits_{j=1}^{i-1}  T_{ij}^{-1} \right)\psi  \ ,
\end{equation}
with $ T_{ij}$ such that
\begin{equation}
\psi^\dagger T_{ij}^{-1} \psi =\int d^M\left( \log\left( \mathbf{y}\right)\right) \ \left|(H[\psi])_{ij}(\mathbf{y})\right|^2 \ .
\end{equation}
Note that this prior is rotationally invariant which indicates that curvature in all directions is treated in the same way.

\subsection{Discrete derivatives}
In order to apply the theoretic discussions above to a finite setting (e.g., a finite grid on a computer) we need a discrete representation of the operators involved in the analysis, in particular of the log-derivative operator. One usual way is to approximate derivatives in terms of finite differences (see e.g., \cite{forsythe1960finite}). A possible way of discretizing the second logarithmic derivative is described in the appendix of \cite{2013PhRvE..87c2136O}.

However, in this particular setting there exist also another way of differentiation in terms of Fourier transformation. We note that, using the chain rule, the second logarithmic derivative of $y$ can be written as
\begin{equation}
	\partial^2_{\log(y)} = y^2 \partial^2_y + y \partial_y \ ,
\end{equation} 
where we restrict our discussion to one dimension, for simplicity.
Furthermore, as discussed in section \ref{sec:sde}, differentiation can be translated to multiplication in harmonic space. I.e.
\begin{equation}
\partial_y \psi(y) = \int d k \ i  k \ \tilde{\psi}(k) \ e^{iky} \ ,
\end{equation}
which indicates that using discrete Fourier transformations, derivatives can be represented by point-wise multiplication in harmonic space.

At first sight, this may seem like a more complicated method for differentiation. However, we note that on a parallel machine, Fourier transformations as well as point-wise multiplications can be fully parallelized while finite differences methods always involve inter-node communication due to subtracting shifted versions of the field representation. On a single node, however, finite differences appear to be computationally more efficient.

Therefore, depending on the problem setting, one method may be superior to the other. Our tests indicate that both methods of differentiation are applicable for prior construction in our problem setting. However, we do not recommend to mix both methods within one inference problem, as the exact form of the derivatives might be incompatible. A more sophisticated test in terms of computational time and accuracy is beyond the scope of this work.

\section{Complex logarithm and smoothness at zero}\label{ap:comlog}
In some applications of the spectral density inference method, we need to impose the smoothness-prior on a zero-centered harmonic space, since the negative part of the density can carry additional information and a shift to purely positive values is not always possible.

As the smoothness-prior involves logarithmic derivatives, we seek to find a way to define logarithmic derivatives for negative values. This is achieved in terms of the complex logarithm. Consider for example $k>0$, then
\begin{equation}
	\log(-k) = \log( e^{i\pi} k)= \log(k) + i\pi \ ,
\end{equation}
and therefore the infinitesimal line-element reads
\begin{equation}
	\left| d\log(-k)\right|  = \left| d\log(k)\right|  \ , \ \forall k \neq 0 \ .
\end{equation}
This indicates that we can express the derivative with respect to a negative $k$ in terms of the corresponding positive differential, i.e.
\begin{equation}
	\left| \frac{\partial \psi(k)}{\partial\log(k)}\right|  = \left| \frac{\partial \psi(k)}{\partial\log(\left| k\right| )}\right|  \ , \ \forall k \neq 0 \ .
\end{equation}
As the logarithm of zero is not defined, the smoothness-prior is also not defined at zero. In this work, we fix this problem by adding a prior to the analysis which aims to minimize the second derivative w.r.t $k$. Therefore, the Hamiltonians of $\tau$ and $\delta$ get modified by a term
\begin{equation}\label{eq:zero}
	H_\eta(\psi) = \frac{1}{2}\psi^\dagger D_\eta^{-1} \psi = \frac{1}{2 \eta^2} \int \left| \frac{\partial^2 \psi(k)}{\partial k^2}\right|^2 dk \ , \ \psi \in \{\tau,\delta\} \ .
\end{equation}
Note that for small $k$ this prior dominates the smoothness prior, while for larger $k$ the logarithmic derivatives are dominant and this second prior does not contribute significantly any more. All applications shown in section \ref{sec:application} use $\eta=0.1$.

\section{Hamiltonian gradients and curvature}\label{ap:grad}
In order to obtain maximum a posterior solutions as well as uncertainty estimates from the information Hamiltonians presented throughout this paper we need to evaluate the corresponding gradients and curvatures.

For the Hamiltonian defined in section \ref{sec:post1}, Eq.\,\ref{eq:h1}, including the modified prior (Eq.\,\ref{eq:zero}) we find
\begin{equation}
	\frac{\partial \mathcal{H}(\tau,\delta)}{\partial \tau_k} =  \frac{1}{2} (1- \phi^*_k \phi_k e^{-\tau_k-\tan(\delta_k)}) + (T_\sigma^{-1} \tau)_k +(D_\eta^{-1} \tau)_k\ ,
\end{equation}
and
\begin{align}
&\frac{\partial \mathcal{H}(\tau,\delta)}{\partial \delta_k} =  \frac{1- \phi^*_k \phi_k e^{-\tau_k-\tan(\delta_k)}}{2 \cos(\delta_k)^2} + \notag\\ &+ \left[ \left( T^{-1}_\tau+D^{-1}_\eta + \frac{1}{\nu^2}\right) \delta\right]_k  \ .
\end{align}

In order go get an estimate of the spectral uncertainty we have to consider the transformed posterior distribution of $\tau$ and $\tan(\delta)$ as they enter the logarithmic spectrum. Specifically,
\begin{equation}
	\mathcal{P}(\tau,\tan(\delta)|\phi) = \mathcal{P}(\tau,\delta|\phi) \left| \frac{\delta \tan(\delta)}{\delta \delta}\right|^{-1} \ ,
\end{equation}
where $\left| \bullet\right| $ denotes the functional determinant.
The corresponding information Hamiltonian reads
\begin{equation}\label{eq:tanham}
	\mathcal{H}(\tau,\tan(\delta)|\phi) = \mathcal{H}(\tau,\delta|\phi) -  \mathrm{Tr}\left( \log\left( \cos(\delta)^2\right) \right) \ .
\end{equation}
The second derivatives of this Hamiltonian can be used to get a Gaussian approximation of the posterior from which we retrieve an uncertainty estimate for the log-spectrum.
The derivatives read
\begin{equation}
\frac{\partial^2 \mathcal{H}(\tau,\tan(\delta))}{\partial \tau_k \ \partial \tau_q} =  \frac{1}{2} \phi^*_k \phi_k e^{-\tau_k-\tan(\delta_k)} \delta_{kq} + (T_\sigma^{-1} +D_\eta^{-1})_{kq}\ ,
\end{equation}
and
\begin{align}
&\frac{\partial^2 \mathcal{H}(\tau,\tan(\delta))}{\partial \tan(\delta_k) \ \partial \tan(\delta_q)} =  \frac{1}{2} \phi^*_k \phi_k e^{-\tau_k-\tan(\delta_k)} \delta_{kq} + \notag\\ &+ \cos^2(\delta_k) \cos^2(\delta_q) \left( T^{-1}_\tau+D^{-1}_\eta + \frac{1}{\nu^2}\right) _{kq} - \notag\\ &- 2 \left( \cos^3(\delta) \sin(\delta) \left( T^{-1}_\tau+D^{-1}_\eta + \frac{1}{\nu^2}\right) \delta -1 \right)_k \delta_{kq} 
 \ .
\end{align}
The square-root of the diagonal of the inverse operators (which we call $\sqrt{\hat{O}}$) can then be regarded as the one-sigma uncertainty estimate of the corresponding quantity. Further details are described in \cite{2013PhRvE..87c2136O}. For the noisy data posterior (Eq.\,\ref{eq:marginalhamilton}) the derivations are completely analogous.
\end{appendix}

\end{document}